\documentclass[preprint,prl,twocolumn,10pt,nobibnotes,byrevtex]{revtex4}%
\usepackage{amsfonts}
\usepackage{amsmath}
\usepackage{amssymb}
\usepackage{graphicx}%
\providecommand{\U}[1]{\protect\rule{.1in}{.1in}}

\begin{document}
\preprint{UATP/1103}
\title{Generalized Non-equilibrium Heat and Work and the Fate of the Clausius Inequality}
\author{P.D. Gujrati}
\email{pdg@uakron.edu}
\affiliation{Department of Physics, Department of Polymer Science, The University of Akron,
Akron, OH 44325}

\begin{abstract}
By generalizing the traditional concept of heat $dQ$ and work $dW$ to also
include their time-dependent irreversible components $d_{\text{i}}Q$ and
$d_{\text{i}}W$ allows us to express them in terms of the instantaneous
internal temperature $T(t)$ and pressure $P(t)$, whereas the conventional form
uses the constant values $T_{0}$ and $P_{0}$ of the medium. This results in an
extremely useful formulation of non-equilibrium thermodynamics so that the
first law turns into the Gibbs fundamental relation and the Clausius
inequality becomes an equality $\oint dQ(t)/T(t)\equiv0$ in \emph{all} cases,
a quite remarkable but unexpected result. We determine the irreversible
components $d_{\text{i}}Q\equiv d_{\text{i}}W$ and discuss how they can be
determined to obtain the generalized $dW(t)$ and $dQ(t)$.

\end{abstract}
\date{\today}
\maketitle

\begin{figure}
[ptb]
\begin{center}
\includegraphics[
trim=0.400466in 0.000000in -0.133316in 0.000000in,
height=1.772in,
width=3.4584in
]%
{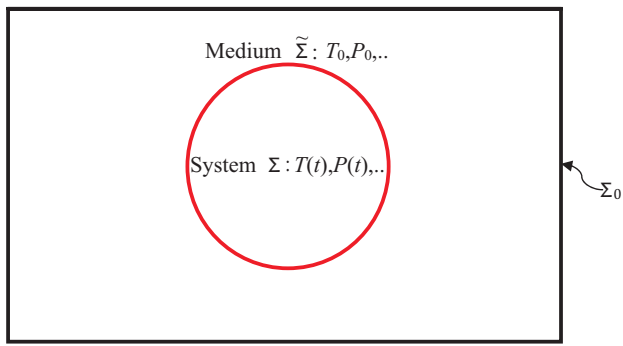}%
\caption{Schematic representation of $\Sigma$, $\widetilde{\Sigma}$ and
$\Sigma_{0}$. We assume that $\Sigma$ and $\widetilde{\Sigma}$ are homogeneous
and in internal equilibrium, but not in equilibrium with each other. The
internal fields $T(t),P(t),\cdots$ fof $\Sigma$ and $T_{0},P_{0},\cdots$ of
$\widetilde{\Sigma}$ are not the same unless they are in equilibrium with each
other. There will be viscous dissipation in $\Sigma$ when not in equilibrium
with $\widetilde{\Sigma}$.}%
\label{Fig_System}%
\end{center}
\end{figure}

Gislason and Craig \cite{Gislason} recently remarked that the definition of
work in non-equilibrium "...thermodynamics processes remains a contentious
topic," a rather surprising statement, as the field of thermodynamics is an
old discipline. However, there is some truth to their critique, which was
motivated by an earlier paper by Bertrand \cite{Bertrand}, who revisited the
confusion first noted by Bauman \cite{Bauman} about different formulation of
work $dW=P_{0}dV$\ or $dW=PdV$ in terms of internal ($P$) and external
($P_{0}$) pressures \cite{note-1}, see Fig. \ref{Fig_System}, and discussed by
many others
\cite{Kievelson,Gislason-TwoSystems,Bizarro,Anacleto-SecondLaw,Anacleto-DissipativeWork,Honig}
since then. Different formulation of work from the first law obviously results
in different heat. We refer the reader to these papers for an interesting
history of the confusion. Gislason and Craig \cite{Gislason-TwoSystems} list
twenty-six representative textbooks including \cite{Kirkwood,Zemansky} where
the pressure-volume work and heat are defined so differently that they are not
equivalent in the presence of irreversibility, and there appears to be no
consensus about their right formulation so far \cite[p.181, Vol. 1]{Kestin}.
There is obviously no problem for reversible processes.%

The first law of thermodynamics, see for example, Landau and Lifshitz
\cite{Landau} or Kondepudi and Prigogine \cite{Prigogine}, relates the change
$dE$ in the internal energy to heat $dQ$ \emph{added to} and the work $dW$
\emph{done by} the system%
\begin{equation}
dE=dQ-dW. \label{First_Law}%
\end{equation}
It is a very general statement and is supposed to be valid for all processes.
We mostly consider mechanical work, but the arguments are valid for all kinds
of work; see near the end. Zemansky \cite[p.73]{Zemansky} defines heat as
energy exchange "...by virtue of a temperature difference only."
Unfortunately, this rules out any isothermal (reversible) heat exchange and
cannot be considered general. Kirkwood and Oppenheim define heat as energy
exchange resulting in "...the temperature increment..." (which rules out phase
changes requiring latent heat) and later note that the work may be converted
to heat due to frictional dissipation \cite[pp.16,17]{Kirkwood} as was first
observed by Count Rumford in 1798 during the boring of cannon \cite{Prigogine}%
. Therefore, it is natural to account for such viscous dissipation in work and
heat when dealing with non-equilibrium systems. The fact that literature is
not very clear on how to incorporate viscous dissipation has motivated this
work; see however \cite{Bizarro,Anacleto-DissipativeWork}, but the authors do
not take the discussion far enough to obtain the results derived here. Our
approach to incorporate viscous dissipation results in expressing the first
law in terms of internal fields and provides an elegant formulation of
non-equilibrium thermodynamics in which the Clausius inequality turns into an
equality in \emph{all} cases, which is a remarkable result in its own right.
It has been recently suggested \cite{Anacleto-SecondLaw} that use of internal
fields is not always consistent with the second law. We find no such problem
in our approach.

We consider our system $\Sigma$ (see Fig. \ref{Fig_System}) surrounded by a
very large medium $\widetilde{\Sigma}$ so large that its fields such as its
temperature $T_{0}$, pressure $P_{0}$, etc.\ are not affected by the system.
They form an isolated system $\Sigma_{0}$.\ In the following, all extensive
quantities pertaining to $\widetilde{\Sigma}$ and $\Sigma_{0}$ carry an
annotation tilde or a suffix $0$, respectively, and those pertaining to
$\Sigma$ carry no suffix. Following modern notation \cite{Prigogine,deGroot},
exchanges with the medium and changes within the system carry the suffix e and
i, respectively.

\paragraph{Traditional Formulation of the First Law}

To truly appreciate our contribution, it is useful to consider how the first
law is traditionally expressed. While the change $dE$ is uniquely defined for
any infinitesimal process, the value of $dQ$ is determined by $dW$. If $dW$
cannot be uniquely determined in irreversible processes because of the
ambiguity noted above, then $dQ$ will be ill-defined. Traditionally, $dQ$
represents the amount of heat exchange $d_{\text{e}}Q$, i.e. the heat given to
the system from the surrounding medium, so that $-dW\ $is identified with the
work exchange $-d_{\text{e}}W\equiv d\widetilde{W}=P_{0}d\widetilde{V}$ by the
medium to the system, giving $d_{\text{e}}W=P_{0}dV$ since $d\widetilde
{V}=-dV$. This is true even if the\emph{ instantaneous} pressure $P$ of the
system is different from $P_{0}$. As the net heat exchange $d_{\text{e}%
}Q+d\widetilde{Q}=0$, we have $d_{\text{e}}Q$ $=-d\widetilde{Q}=-T_{0}%
d\widetilde{S}=T_{0}d_{\text{e}}S$ in terms of the entropy change
$d_{\text{e}}S$ of the system. In other words,
\begin{equation}
dE=d_{\text{e}}Q-d_{\text{e}}W=T_{0}d_{\text{e}}S-P_{0}%
dV\label{Standard_Heat_Sum}%
\end{equation}
expressed in terms of either exchange quantities or external fields of
$\widetilde{\Sigma}$. This is Protocol 3 of Bertrand \cite{Bertrand}.

However, the situation is not always clear. Kondepudi and Prigogine introduce
the work by equating it to $PdV$, where $P$ "...is the pressure at the moving
surface," but they do not mention whether the form is applicable to all
processes. Landau and Lifshitz are explicit and state that $dW=PdV$ for
reversible and irreversible processes \cite[p.45]{Landau}. They require for
this the \emph{existence} of mechanical equilibrium (and so do
\cite{Kirkwood,Zemansky}) within $\Sigma$ so that at each instant during the
process $P$ must be uniform throughout the body; its equality with $P_{0}$ is
not required. However, they do not discuss $dQ$ when they consider $\Sigma$
and $\widetilde{\Sigma}$ out of equilibrium \cite[Sect. 20]{Landau}. If we use
$dW=PdV$ for the work \emph{done by} $\Sigma$\ in terms of $P$ of the system
(this is similar to Protocol 4 of Bertrand \cite{Bertrand}), then this will
alter the heat $dQ$ \emph{added to} $\Sigma$ in Eq. (\ref{First_Law}) for $dE$
must be invariant to the choice of internal or external fields. To the best of
our knowledge, this issue of the actual forms of $dQ$ in the two protocols and
what is the correct form of $dW$ for non-equilibrium systems has not been
discussed in the literature, even though Kestin \cite[Sect. 5.12]{Kestin}
clearly states that distinguishing heat and work in non-equilibrium states is
not unambiguous.

\paragraph{General Consideration and\ Clausius Relation}

The first law is not useful for any computation unless we can ascribe
temperatures, pressures, etc. to $\Sigma$ and $\widetilde{\Sigma}$. This is
done by taking them to be in \emph{internal equilibrium}
\cite{Landau,Gujrati-I} so that their instantaneous entropies are \emph{state
functions} $S(t)=S(E(t),V(t),\cdots)$ and $\widetilde{S}(t)=S(\widetilde
{E}(t),\widetilde{V}(t),\cdots)$ of (time-dependent) state variables. They
change as respective state variables $E(t),V(t),\cdots$ or $\widetilde
{E}(t),\widetilde{V}(t),\cdots$ change with time. We first only consider
energy and volume. The temperatures and pressures are given by appropriate
standard derivatives of the entropies; see Eq.
(\ref{Gibbs_Fundamental_Relations}). It now follows that the Gibbs fundamental
relations are given by \cite{Landau}%
\begin{equation}
dE=TdS-PdV,\ \ d\widetilde{E}=T_{0}d\widetilde{S}-P_{0}d\widetilde{V};
\label{Gibbs_Fundamental_Relations}%
\end{equation}
quantities without the suffix $0$ normally depend on time. The validity of Eq.
(\ref{Gibbs_Fundamental_Relations}) requires $\Sigma$ and $\widetilde{\Sigma}$
to be independently homogeneous \cite{Landau,deGroot,Gujrati-I}. We
have\ $dE_{0}=dV_{0}=0$ for $\Sigma_{0}$. The application of the first law
gives%
\[
dE=dQ-dW,\ d\widetilde{E}=d\widetilde{Q}-d\widetilde{W},\ dE_{0}=dQ_{0}%
-dW_{0}=0.
\]
We take $dQ=d_{\text{e}}Q+d_{\text{i}}Q$ as a generalization of
$dQ=d_{\text{e}}Q$ in Eq. (\ref{Standard_Heat_Sum}): it denotes the heat added
to the system either through exchange with its exterior ($d_{\text{e}}Q$) or
by dissipative internal forces within ($d_{\text{i}}Q$). Similarly,
$dW=d_{\text{e}}W+$ $d_{\text{i}}W=PdV$ is the work done by the system on its
exterior ($d_{\text{e}}W$) and by dissipative internal forces ($d_{\text{i}}%
W$). A somewhat similar looking approach, but different in spirit, is taken in
\cite{Honig-0}.

To make the our approach computationally useful, we need to determine
$d_{\text{i}}Q$\ and $d_{\text{i}}W$; see later. With our generalization,
$dQ_{0}\equiv d_{\text{i}}Q_{0}$ and $dW_{0}\equiv d_{\text{i}}W_{0}$ and
$dQ_{0}\equiv dW_{0}$ for $\Sigma_{0}$. Let us assume that $P>P_{0}$. The
(internal) work done by the pressure difference $\Delta P=P-P_{0}>0$ is
$dW_{0}=\Delta PdV>0$ , since $dV>0$. This results in raising the kinetic
energy $dK_{\text{S}}$ of the center-of-mass of the surface separating
$\Sigma$ and $\widetilde{\Sigma}$ and overcoming work $dW_{\text{fr}}$ done by
all sorts of viscous or frictional drag. Thus,
\[
dW_{0}\equiv d_{\text{i}}W_{0}\equiv dK_{\text{S}}+dW_{\text{fr}}.
\]
Because of the stochasticity associated with any statistical system, both
energies on the right side dissipate among the particles so as to increase the
entropy and appear in the form of heat ($=d_{\text{i}}Q_{0}>0$) within the
isolated system. This is the generalized heat $dQ_{0}>0$, since $d_{\text{e}%
}Q_{0}=0$ for $\Sigma_{0}$. It is normal to associate all irreversible
components in $\Sigma_{0}$\ with $\Sigma$ ($d_{\text{i}}Q\equiv d_{\text{i}%
}Q_{0},d_{\text{i}}W\equiv d_{\text{i}}W_{0}$) and not with the (extensively
large) medium $\widetilde{\Sigma}$, which always remain in internal
equilibrium \cite{Gujrati-I}; compare with the situation of finite
surroundings considered by Bizarro \cite{Bizarro}. Thus, when there are
irreversible processes going on, it is natural to generalize heat from
$d_{\text{e}}Q\ $in Eq. (\ref{Standard_Heat_Sum}) to include the irreversible
heat $d_{\text{i}}Q$ and identify $dQ$ as the heat \emph{added to the system}.
Similarly, we need to generalize work from $d_{\text{e}}W=P_{0}dV$ to $dW=PdV$
and identify it as work \emph{done by the system}. Remarkably, we find that
\begin{equation}
d_{\text{i}}Q\equiv d_{\text{i}}W.\label{Irreversible_Heat_Work_equality}%
\end{equation}
The generalization does not change $dE$ and we have
\begin{equation}
dE=d_{\text{e}}Q-d_{\text{e}}W\equiv dQ-dW=dQ-PdV.\label{General_First_Law}%
\end{equation}
However, the most important result is that \cite{Gujrati-I}
\begin{equation}
dQ=TdS,\label{Clausius_Equality_Differential}%
\end{equation}
even when the system is \emph{not} in equilibrium with the medium. This is
easily seen by comparing Eq. (\ref{General_First_Law}) with the Gibbs
fundamental relation in Eq. (\ref{Gibbs_Fundamental_Relations}). This makes
$dQ/T$ an \emph{exact differential }$dS$ so that
\begin{equation}
\oint dQ/T\equiv0\ \ \text{for}\ \text{any cyclic process.}%
\label{Clausius_Equality_0}%
\end{equation}
This leads to%
\begin{equation}%
{\textstyle\oint}
d_{\text{i}}Q/T=-%
{\textstyle\oint}
d_{\text{e}}Q/T\geq0,\label{Clausius_Inequality}%
\end{equation}
which is consistent with the traditional Clausius inequality $%
{\textstyle\oint}
d_{\text{e}}Q/T$ $\leq0$. Thus, the Clausius equality in Eq.
(\ref{Clausius_Equality_0}) should not be interpreted as the absence of
irreversibility, as is clear from the above inequality. It is only because of
the use of the generalized heat $dQ$ that this inequality has become an
equality. The second integral with its sign in Eq. (\ref{Clausius_Inequality})
can be thought of as the uncompensated transformation $N$ \cite{Prigogine}.
Our formulation has allowed us to identify it with the first integral%
\begin{equation}
N\equiv%
{\textstyle\oint}
d_{\text{i}}Q/T=%
{\textstyle\oint}
d_{\text{i}}W_{0}/T\text{,}\label{Uncompensated_Transformation}%
\end{equation}
which provides a way of computing $N$ using equations of state of $\Sigma$ and
$\widetilde{\Sigma}$. From $d_{\text{i}}S\equiv dS-d_{\text{e}}S$, we find
that
\[%
{\textstyle\oint}
d_{\text{i}}S=-%
{\textstyle\oint}
d_{\text{e}}Q/T_{0}\geq0,
\]
which is the second law for a cyclic process and represents the irreversible
entropy generated in a cycle; recall that $d_{\text{e}}Q$ is the traditional
$dQ$ in Eq. (\ref{Standard_Heat_Sum}).

\paragraph{General Form of Work}

Let us follow the consequences of this particular extension a bit further and
prove that only $dW=PdV$ is consistent with the second law. Since
$dQ_{0}=dQ+d\widetilde{Q}=dW_{0}=dW+d\widetilde{W}\geq0$, we find that the net
heat $dQ_{0}$ in the medium and the system need not vanish, in contrast to
what happens in equilibrium:%
\begin{equation}
dQ+d\widetilde{Q}=d_{\text{i}}W_{0}\geq0 \label{Heat_Sum}%
\end{equation}
because of viscous dissipation $d_{\text{i}}Q_{0}\equiv d_{\text{i}}W_{0}$,
and is related to the irreversible entropy generation within $\Sigma_{0}$ or
$\Sigma$, as we will see later. For frictional forces only ($P=P_{0}$) which
are included in our approach \cite{Gujrati-II}, the above relation reduces to
Eq. (7) of Bizarro \cite{Bizarro}. From $dW+d\widetilde{W}=PdV+P_{0}%
d\widetilde{V}$ valid for any arbitrary $dV$, it follows that%
\begin{equation}
dW=PdV,\ \ \text{ }d\widetilde{W}=P_{0}d\widetilde{V}, \label{Work_Forms}%
\end{equation}
which proves the above assertion. The alternate choice $dW=P_{0}dV$ and
$d\widetilde{W}=Pd\widetilde{V}$ employed in the traditional formulation, see
Eq. (\ref{Standard_Heat_Sum}), will violate the second law as the choice will
lead to negative heat dissipation $dQ_{0}=dW_{0}=(P_{0}-P)dV<0$, a physical
impossibility. Thus, we must write the first law for the system and the
medium, respectively, as
\begin{equation}
dE=dQ-PdV,\ d\widetilde{E}=d\widetilde{Q}-P_{0}d\widetilde{V}.
\label{First_Law_Volume}%
\end{equation}

The above demonstration establishes that the work is given by Eq.
(\ref{Work_Forms}) in terms of the instantaneous pressure in \emph{all} cases
contrary to the traditional wisdom, see Eq, (\ref{Standard_Heat_Sum}), that it
is given by $P_{0}dV$. The generalized formulation brings out a symmetry in
the form of $dW$ and $d\widetilde{W}$\ under the interchange
system$\Longleftrightarrow$medium: the work for any body is always uses its
own internal pressure. This symmetry is absent in the traditional formalism in
which both terms are given by the external pressure $P_{0}$.

\paragraph{Benefits of the Generalized Formulation}

The new formulation has many other desirable properties. For example, the
entire thermodynamics and stability considerations for any body can be
expressed in terms of the variables associated with the body alone at each
instant. We can use the equations of state of the body alone for thermodynamic
computation. The first law is no longer different from the Gibbs fundamental
relation so we only deal with equalities and not inequalities. The generalized
heat $dQ$\ and work $dW$ differ from $d_{\text{e}}Q$ and $d_{\text{e}}W$,
respectively, by the same contribution $d_{\text{i}}W\equiv d_{\text{i}}W_{0}%
$; the latter is the term $d\Sigma$ in Eq. (19) of Eu \cite{Eu}, which can be
absorbed in his work term ($\equiv-d_{\text{e}}W$ in our notation) to reduce
this equation to the above form for $dE$. The determination of $d_{\text{i}%
}W_{0}$ is straight forward by measuring $P$ and $P_{0}$. Adding this to
$d_{\text{e}}Q$ then allows us to determine $dQ$. Consider, as an example, the
case when the medium pressure $P_{0}=0$ such as when a gas expands in vacuum.
In this case, $d_{\text{e}}W=0$; $dW$ is the irreversible work, which cannot
be less than zero, as it is given by the irreversible entropy generation as
shown below.

\paragraph{Irreversible entropy generation in an isothermal process}

We now give an alternative derivation of Eq. (\ref{Work_Forms}) by computing
the irreversible entropy gain so that the role of the latter can be clearly
seen in an irreversible process. A brief discussion of this issue can be found
in Kondepudi and Prigogine \cite[pp. 94-95]{Prigogine} where they discuss
irreversible expansion of a gas; see also \cite{Gujrati-I}. We will assume
that the temperature of the system and the medium is constant during the
process at $T_{0}$. The pressure difference gives rise to dissipation, which
results in an irreversible heat generation $d_{\text{i}}Q=T_{0}d_{\text{i}}S$
in addition to the heat $d_{\text{e}}Q=-d\widetilde{Q}=-T_{0}d\widetilde
{S}=T_{0}d_{\text{e}}S$ received from the medium; here, $d_{\text{i}%
}S=d_{\text{i}}S^{\text{(V)}}$ is the irreversible entropy gain
\cite{Prigogine,Gujrati-I}%
\[
d_{\text{i}}S^{\text{(V)}}=(P-P_{0})dV/T_{0}\geq0.
\]
We see that $dQ=T_{0}(d_{\text{e}}S+d_{\text{i}}S)=T_{0}dS$ in accordance with
Eq. (\ref{Clausius_Equality_Differential}) so that by including $d_{\text{i}%
}Q$ in $dQ$ relates $dQ$ to the entropy change $dS$. The generalized first law
for the medium in Eq. (\ref{First_Law_Volume}) reads $d\widetilde{E}%
=T_{0}d\widetilde{S}-P_{0}d\widetilde{V}$, in which we can replace
$d\widetilde{V}$ by $-dV$ and $d\widetilde{E}$ by $-dE$ to obtain
$dE=-d\widetilde{Q}-P_{0}dV.$ We compare this with the right side of Eq.
(\ref{First_Law}) for the system and use the fact that
\[
dQ+d\widetilde{Q}=d_{\text{i}}Q_{0}=d_{\text{i}}Q
\]
to determine the work $dW$. Simple algebra immediately leads to $dW=PdV,$ as
claimed above. We also note that we do not need Jarzynski result \cite{Honig}
to obtain $d_{\text{i}}S^{\text{(V)}}$.

\paragraph{General Irreversible Process}

We now consider the same process as above, except that the temperature $T$ of
the system is also allowed to be different from that of the medium, which we
take to be the constant $T_{0}$. The general relation in Eq.
(\ref{Irreversible_Heat_Work_equality}) is still valid. However, $d_{\text{i}%
}S$ consists of $d_{\text{i}}S^{\text{(V)}}$ and $d_{\text{i}}S^{\text{(Q)}}$
or $d_{\text{i}}S^{\text{(S)}}$; the former is given above and the latter
contribution is given by%
\[
d_{\text{i}}S^{\text{(Q)}}=(1/T-1/T_{0})dQ=-(T-T_{0})dS/T_{0}=d_{\text{i}%
}S^{\text{(S)}}\geq0.
\]
We immediately find that $d_{\text{i}}S^{\text{(S)}}$ and $d_{\text{i}%
}S^{\text{(V)}}$ are formally of the same form; see \cite{Gujrati-I}. To be
convinced of the form of $d_{\text{i}}S^{\text{(Q)}}\equiv d_{\text{i}%
}S^{\text{(S)}}$, we start with the identity $d_{\text{e}}Q-P_{0}dV\equiv
TdS-PdV$ , which can be transformed into $T_{0}d_{\text{i}}S=(T_{0}%
-T)dS+(P-P_{0})dV$ \cite{Gujrati-I}. We\ can now identify the above two
components of $d_{\text{i}}S$. To obtain $d_{\text{i}}S^{\text{(Q)}}$, we use
Eq. (\ref{Clausius_Equality_Differential}) to replace $dS$. Both irreversible
components are non-negative as expected from the second law. We thus see that
expressing the first law in terms of internal fields is consistent with
irreversible entropy production, contrary to a recent claim
\cite{Anacleto-SecondLaw}.

Both irreversible entropy gains can be easily determined. The determination
requires measuring the two temperatures and pressures, while the determination
of $dQ$ has already been explained. Thus, the generalized formulation creates
no additional problem in experimentally determining various quantities.

\paragraph{Inclusion of other state variables}

Let us now extend the discussion to include other extensive quantities such as
the flow of matter, the electric interactions, chemical reactions, etc. For
specificity, we focus on chemical reactions, which we assume to be described
by a single extent of reaction $\xi$. The corresponding affinity for the
system is given by $A$, while that for the medium is given by $A_{0}=0$. We
assume another observable $X$ such as the number of solvent in a binary
mixture. The corresponding chemical potential is $\mu$ for the system and
$\mu_{0}$\ for the medium. The work is now $dW=PdV-\mu dX+Ad\xi$
\cite{Alberty}. The Gibbs fundamental relation for $\Sigma$ is given by
$dE=TdS-PdV+\mu dX-Ad\xi$, while the first law for it takes the form
$dE=dQ-PdV+\mu dX-Ad\xi$. Comparing the two we find that Eq.
(\ref{Clausius_Equality_Differential}) still holds so that the validity of Eq.
(\ref{Clausius_Equality_0}) is not affected by the presence of other variables
in the first law \cite{Gujrati-II}. Rewriting $dQ=dE+PdV-\mu dX+Ad\xi$ as
$dQ=dE+P_{0}dV-\mu_{0}dX+(P-P_{0})dV-(\mu-\mu_{0})dX+Ad\xi,$ we can identify
\cite{Gujrati-II} $d_{\text{e}}Q$ with the first three terms in which
$d_{\text{e}}W=P_{0}dV-\mu_{0}dX$ for work of all types \cite{Alberty}.
Similarly, $d_{\text{i}}Q=d_{\text{i}}W\equiv dW_{0}$ represents the last
three terms $(P-P_{0})dV-(\mu-\mu_{0})dX+Ad\xi$; see Eq.
(\ref{Irreversible_Heat_Work_equality})$.$ We again obtain the general results
in Eq. (\ref{General_First_Law}), as expected. According to the second law,
each contribution in $d_{\text{i}}W$ is non-negative. The extension to
arbitrary number of observables and internal variables is trivial
\cite{Gujrati-II}.

In summary, we have shown that the definition of the generalized heat $dQ$,
which includes its irreversible component $d_{\text{i}}Q$, follows uniquely
from the unique choice of $dW$ resulting from the second law; see Eq.
(\ref{Work_Forms}). It includes its irreversible component $d_{\text{i}%
}W\equiv$ $d_{\text{i}}Q$; see Eq. (\ref{Irreversible_Heat_Work_equality}).
Another remarkable consequence is that in terms of generalized $dQ$,\ the
Clausius equality (\ref{Clausius_Equality_0}) is always maintained, in
contrast to the inequality (\ref{Clausius_Inequality}) in the traditional
approach. Our generalized formulation brings about a symmetry between the
system and its surrounding medium, see Eq. (\ref{Work_Forms}), which is absent
in the traditional approach using external fields. The first law becomes
identical with the Gibbs fundamental relation so that we only deal with
equalities that are easier to deal with than the inequalities in the
traditional approach.

\end{document}